\documentclass[conference]{IEEEtran}

\usepackage[utf8]{inputenc}

\usepackage[nobreak]{cite}
\usepackage{amsmath,amssymb,amsfonts}
\usepackage{graphicx}
\usepackage{textcomp}
\usepackage{xcolor}

\usepackage{algorithm}
\usepackage[noend]{algpseudocode}

\algrenewcommand\textproc{} %
\newenvironment{ls_algorithmic}
{\begin{algorithmic}[1]\small}
    {\end{algorithmic}}
\algrenewcommand\algorithmicindent{1.0em}
\algtext*{EndIf}
\algtext*{EndFor}
\algtext*{EndWhile}
\algtext*{EndFunction}
\algnewcommand\algorithmicforeach{\textbf{for each}}
\algdef{S}[FOR]{ForEach}[1]{\algorithmicforeach\ #1\ \algorithmicdo}
\newcommand\CONDITION[2]%
{\begin{tabular}[t]{@{}l@{}l@{}}
    #1 & #2
  \end{tabular}%
}
\algdef{S}[IF]{IfLongCond}[1]%
{\algorithmicif\ \CONDITION{#1}{\ \algorithmicthen}}%

\definecolor{custom_red}{rgb}{0.7,0,0}
\newcommand*{\tr}[1]{\textrm{#1}}
\newcommand*{\ti}[1]{\textit{#1}}

\newcommand*{\fgpdg}[3]{$\tr{G}(\tr{#1}^\tr{#2}_\tr{#3})$}
\newcommand*{\astree}[3]{$\tr{A}(\tr{#1}^\tr{#2}_\tr{#3})$} %
\newcommand*{\code}[3]{$\tr{#1}^\tr{#2}_\tr{#3}$}
\newcommand*{\mapping}[2]{$\tr{M}^\tr{#1}_\tr{#2}$}

\newtheorem{df}{Definition}

\usepackage{url}
\usepackage[hidelinks]{hyperref}

\usepackage{mdframed}
\mdfdefinestyle{light-blue}{%
backgroundcolor=BGLightBlue,
innertopmargin=2,
innerbottommargin=2,
innerleftmargin=4,
innerrightmargin=4,
outerlinewidth=0pt,
topline=false,
rightline=false,
leftline=false,
bottomline=false,
linewidth=1pt,
skipabove=3,
skipbelow=3,
}

\mdfdefinestyle{simple-frame}{%
innertopmargin=2,
innerbottommargin=2,
innerleftmargin=4,
innerrightmargin=4,
outerlinewidth=0pt,
topline=true,
rightline=true,
leftline=true,
bottomline=true,
linewidth=.5pt,
skipabove=3,
skipbelow=2,
}

\usepackage{xpatch}
\makeatletter
\xpatchcmd{\endmdframed}
  {\aftergroup\endmdf@trivlist\color@endgroup}
  {\endmdf@trivlist\color@endgroup\@doendpe}
  {}{}
\makeatother

\usepackage{cleveref}
\Crefname{figure}{\figurename}{Figs.}
\Crefname{lstlisting}{Listing}{Listings}
\usepackage{myref_en}

\newcommand*{\Sprecision}{0.710}
\newcommand*{\Srecall}{0.565}
\newcommand*{\Sfscore}{0.630}

\begin{document}

\title{Sirius: Static Program Repair with Dependence Graph-Based Systematic Edit Patterns}

\author{\IEEEauthorblockN{Kunihiro Noda, Haruki Yokoyama, and Shinji Kikuchi}
  \IEEEauthorblockA{\textit{Fujitsu Research, Japan}\\
    \{noda.kunihiro, yokoyama.haruki, skikuchi\}@fujitsu.com}
}

\maketitle

\begin{abstract}

Software development often involves \textit{systematic edits}, similar but nonidentical changes to many code locations, that are error-prone and laborious for developers.
Mining and learning such systematic edit patterns (SEPs) from past code changes enable us to detect and repair overlooked buggy code that requires systematic edits.

A recent study presented a promising SEP mining technique that is based on program dependence graphs (PDGs), while traditional approaches leverage syntax-based representations.
PDG-based SEPs are highly expressive and can capture more meaningful changes than syntax-based ones.
The next challenge to tackle is to apply the same code changes as in PDG-based SEPs to other code locations;
 detection and repair of overlooked locations that require systematic edits.
Existing program transformation techniques cannot well address this challenge
 because (1) they expect many structural code similarities that are not guaranteed in PDG-based SEPs
 or (2) they work on the basis of PDGs but are limited to specific domains (e.g., API migrations).

We present in this paper a general-purpose program transformation algorithm for applying PDG-based SEPs.
Our algorithm identifies a small transplantable structural subtree for each PDG node, thereby adapting code changes from PDG-based SEPs to other locations.
We construct a program repair pipeline \textit{Sirius} that incorporates the algorithm and automates the processes of mining SEPs, detecting overlooked code locations (bugs) that require systematic edits, and repairing them by applying SEPs.

We evaluated the repair performance of Sirius with a corpus of open source software consisting of over 80 repositories.
The results indicate that Sirius greatly outperformed the state-of-the-art technique for syntax-based SEPs.
Sirius achieved a precision of \Sprecision{}, recall of \Srecall{}, and F1-score of \Sfscore{}, while those of the state-of-the-art technique were 0.470, 0.141, and 0.216, respectively.

\end{abstract}

\begin{IEEEkeywords}
  automated program repair, program dependence graph, systematic edit patterns, program transformation
\end{IEEEkeywords}

\section{Introduction}
\label{sec:introduction}

The repetitiveness of code changes has been empirically observed\tcite{systematic-code-changes:icse09,repetitive-code-changes:ase2013,plastic-surgery:fse:2014}.
Similar but nonidentical changes to many code locations, called \textit{systematic edits}, are often required in several development activities such as continual feature additions, bug fixes, and API migrations.
Manually searching and applying nonidentical similar edits to many locations are quite error-prone, and developers sometimes overlook some code locations that require systematic edits.

By capturing such repetitive edits in reusable forms as systematic edit patterns (SEPs), we can automate the error-prone tasks and detect/repair the overlooked code locations to edit.
Various studies on SEPs have been conducted for assisting development activities:
  mining SEPs from change deltas in development histories\tcite{cpatminer:icse2019,sysedminer:msr2017,c3:msr2016,ammonia:emse2020,sbd:fse2013};
  learning systematic program transformations from change examples and applying them elsewhere\tcite{genpat:ase2019,refazer:icse2017,ares:fse2017,lase:icse2013,sydit:pldi2011,recommend-edit-locations:saner2017,lac:compsac2019};
  building repair pipelines that automate the processes of mining and applying SEPs\tcite{precfix:icse2020}.

To mine SEPs, most techniques take syntax-based (token/AST diff-based) approaches that capture similar structural changes and generalize them to patterns\tcite{c3:msr2016,sysedminer:msr2017,change-mining-in-ide:icse2014}.
Recently, a dependence graph-based SEP miner, called CPatMiner, has been presented\tcite{cpatminer:icse2019}.
It represents code edits as changes of program dependence graphs (PDGs) and detects frequent and common graph patterns, resulting in PDG-based SEPs.
Semantic change patterns (PDG-based SEPs) obtained from CPatMiner are more complex and meaningful than SEPs mined using syntax-based techniques.

The next important challenge to tackle is to apply the same changes as in PDG-based SEPs to other code locations,
 meaning detection and repair of overlooked locations that require systematic edits.
Most existing program transformation techniques cannot effectively address this challenge
 because they expect many structural similarities among pattern instances and locations where SEPs are applied.
Such structural similarities are not guaranteed when using PDG-based SEPs.
There are a few transformation techniques based on semantic edit patterns\tcite{a3:tse2020,CocciEvolve:icpc2020,libsync:oopsla2010,phoenix:fse2019},
 but they are all limited to specific domains (e.g., API migrations).
There are no general-purpose transformation techniques for PDG-based SEPs presented yet.

In this paper, we propose a general-purpose program transformation algorithm for applying PDG-based SEPs to other code locations.
We build a program repair pipeline \textit{Sirius} that incorporates our algorithm.
It automates the processes of mining PDG-based SEPs from repositories and detecting/repairing overlooked code locations that require systematic edits.
In Sirius, we leverage PDG-based representation of SEPs used in CPatMiner\tcite{cpatminer:icse2019} with slight modifications, in which an SEP is represented as a change from old to new PDGs.

Sirius consists of three components: a miner, detector, and transformer.
The miner extracts past code changes as PDG changes, then obtains PDG-based SEPs by frequent subgraph mining.
The detector component detects locations where SEPs can be applied in the given client code (repair targets).
It represents client code as PDGs and finds matches against SEP graphs; matched locations (subgraphs) can be transformed according to SEPs.
The transformer applies the same changes as in SEPs to client locations identified by the detector.

One straightforward idea for applying a PDG-based SEP is to transform the client PDG such that it contains the same structure as the new PDG of the SEP.
However, this approach is infeasible because it is impossible to reconstruct program code (esp., control flows) from the partially transformed PDG
(as shown in \Cref{sec:motivating-example}).
To overcome this difficulty, instead of directly transforming the client PDG,
 our algorithm identifies a transplantable induced subtree of the AST for each node in a PDG-based SEP, then adapts it to the client AST.

We evaluated the effectiveness of Sirius with a corpus of open source software (OSS).
By mining the corpus consisting of 80+ real-world projects, we obtained 6,060 PDG-based SEPs.
With the mined patterns, we conducted cross-validation to test whether Sirius can correctly simulate (i.e., learn and apply) a systematic edit for each pattern instance,
 which involved 31,273 repair trials.
As a result, 
Sirius achieved a precision of \Sprecision{}, recall of \Srecall{}, and F1-score of \Sfscore{} while those of the state-of-the-art technique were 0.470, 0.141, and 0.216, respectively.
The contributions of this paper are:
\begin{itemize}
  \item a novel general-purpose program transformation algorithm for PDG-based SEPs;
  \item a program repair pipeline \textit{Sirius} that fully automates the processes of mining SEPs, detecting overlooked code locations (bugs) that require systematic edits, and repairing them by applying SEPs;
  \item a quantitative evaluation with a corpus of 80+ real-world projects, which shows the effectiveness of Sirius.
\end{itemize}

\section{Motivating Example}
\label{sec:motivating-example}

\begin{figure}[tb]
  \centering
  \includegraphics[width=.9\columnwidth]{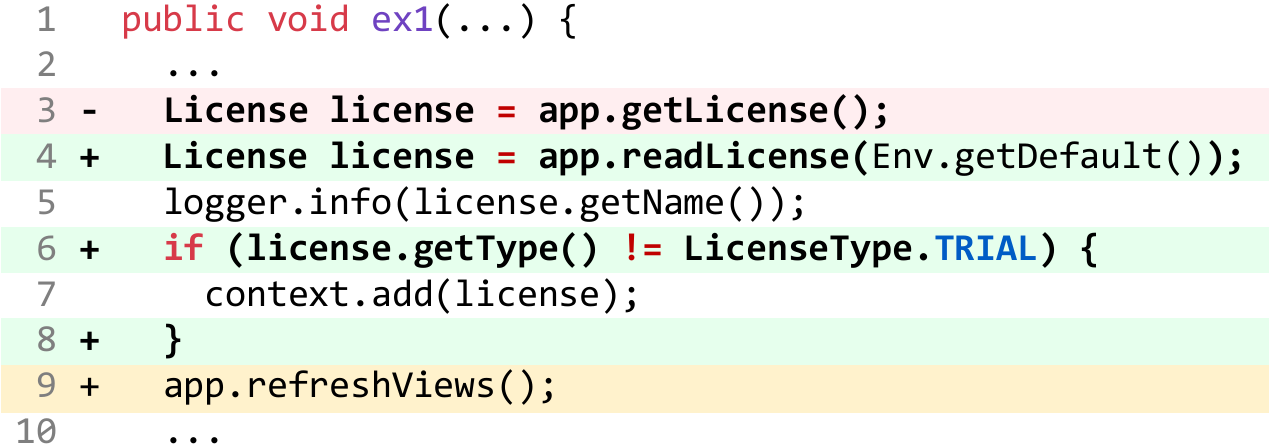}
  \caption{Code-change example 1 ($\tr{W}_{\tr{ex1}}$).}
  \label{fig:code-delta1}
\end{figure}

\begin{figure}[tb]
  \centering
  \includegraphics[width=.9\columnwidth]{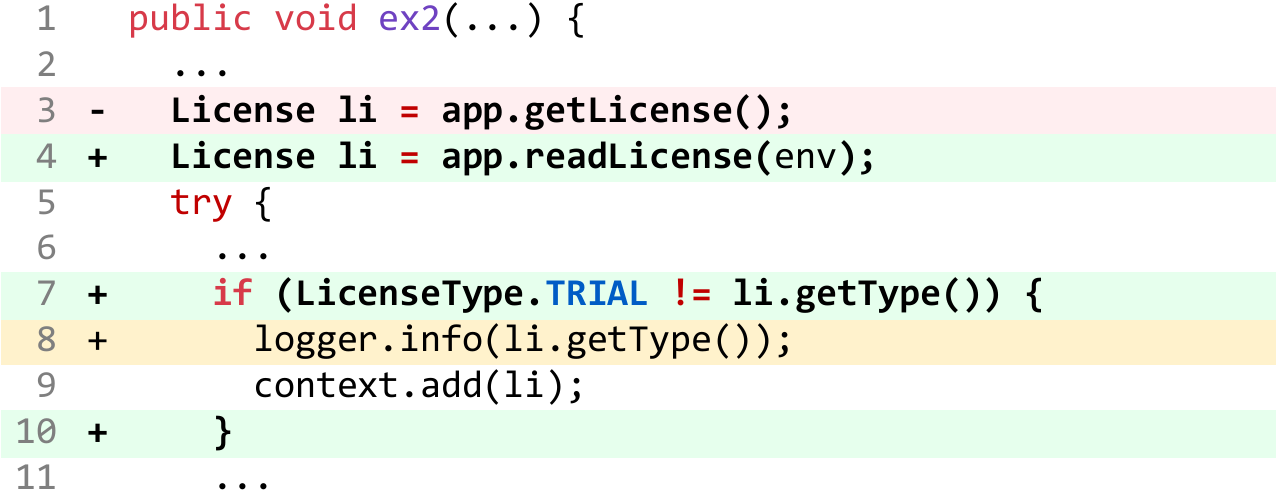}
  \caption{Code-change example 2 ($\tr{W}_\tr{ex2}$).}
  \label{fig:code-delta2}
\end{figure}

We take the two past code changes shown in \Cref{fig:code-delta1,fig:code-delta2} as our motivating example.
The changes have a common edit pattern: a deprecated method \textit{getLicense} is replaced with a newer one \textit{readLicense}, and \textit{context.add} is surrounded with an \textit{if} statement regarding license types.
There are also irrelevant edits (colored in yellow); they will not be included in a mined SEP because they are not common between the two changes.

\begin{figure}[tb]
  \begin{minipage}{.58\columnwidth}
    \centering
    \includegraphics[width=.95\columnwidth]{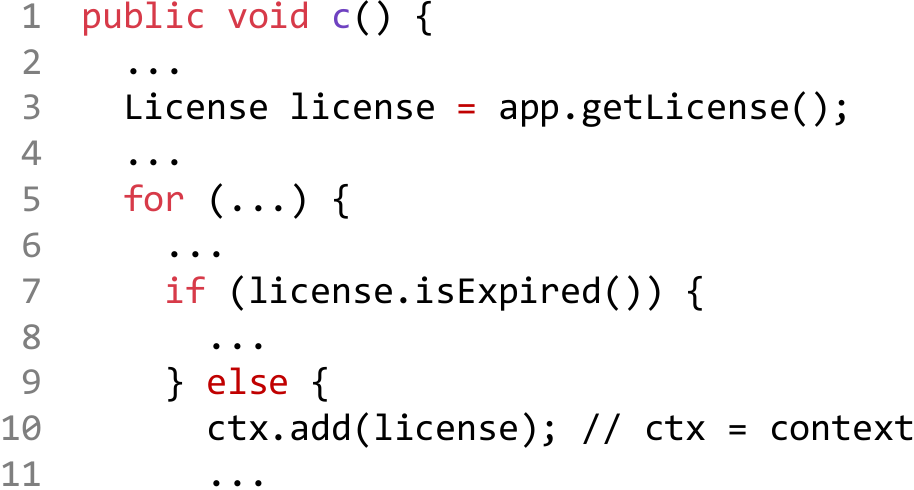}
    \caption{Client method (\code{W}{}{c}).}
    \label{fig:client-method}
  \end{minipage}
  \begin{minipage}{.41\columnwidth}
    \centering
    \includegraphics[width=\columnwidth]{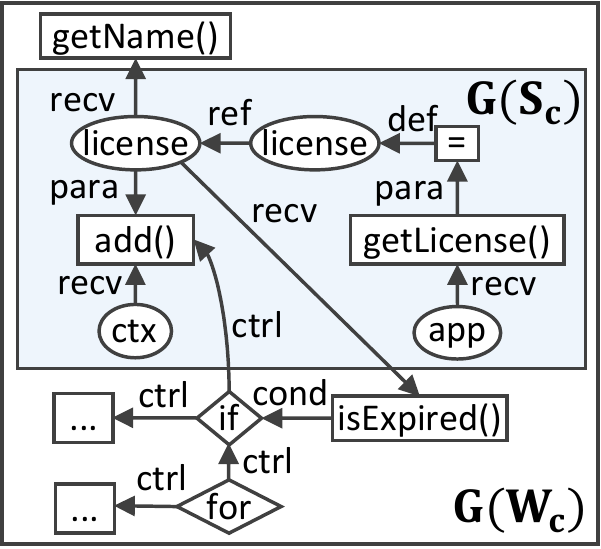}
    \caption{fgPDG of \code{W}{}{c}.}
    \label{fig:client-fgpdg}
  \end{minipage}
\end{figure}

We assume that a client method (\Cref{fig:client-method}) that still uses the deprecated \textit{getLicense} exists at the latest revision of the client project.
What we would like to do with \textit{Sirius} are to extract an SEP (how to change code) from the past changes (\Cref{fig:code-delta1,fig:code-delta2}),
 detect the overlooked location that requires the systematic edit (l.3 and l.10 in \Cref{fig:client-method}),
 and repair it based on the SEP.

\subsection{Change Pattern Representation with Fine-Grained PDGs}
\label{sec:change-graph-representation}

We exploit the PDG-based representation of SEPs used in CPatMiner\tcite{cpatminer:icse2019}.
The program code of a method is represented as a \textit{fine-grained PDG} (fgPDG), and a code change is represented as a \textit{change graph}.
An fgPDG captures data/control dependencies at an expression level and expresses API and object usages on the basis of GROUM\tcite{groum:fse2009}.
A change graph contrasts the old and new fgPDGs (i.e., fgPDGs before and after code changes) by placing the two graphs side by side.%

\begin{figure}[tb]
  \centering
  \includegraphics[width=.95\columnwidth]{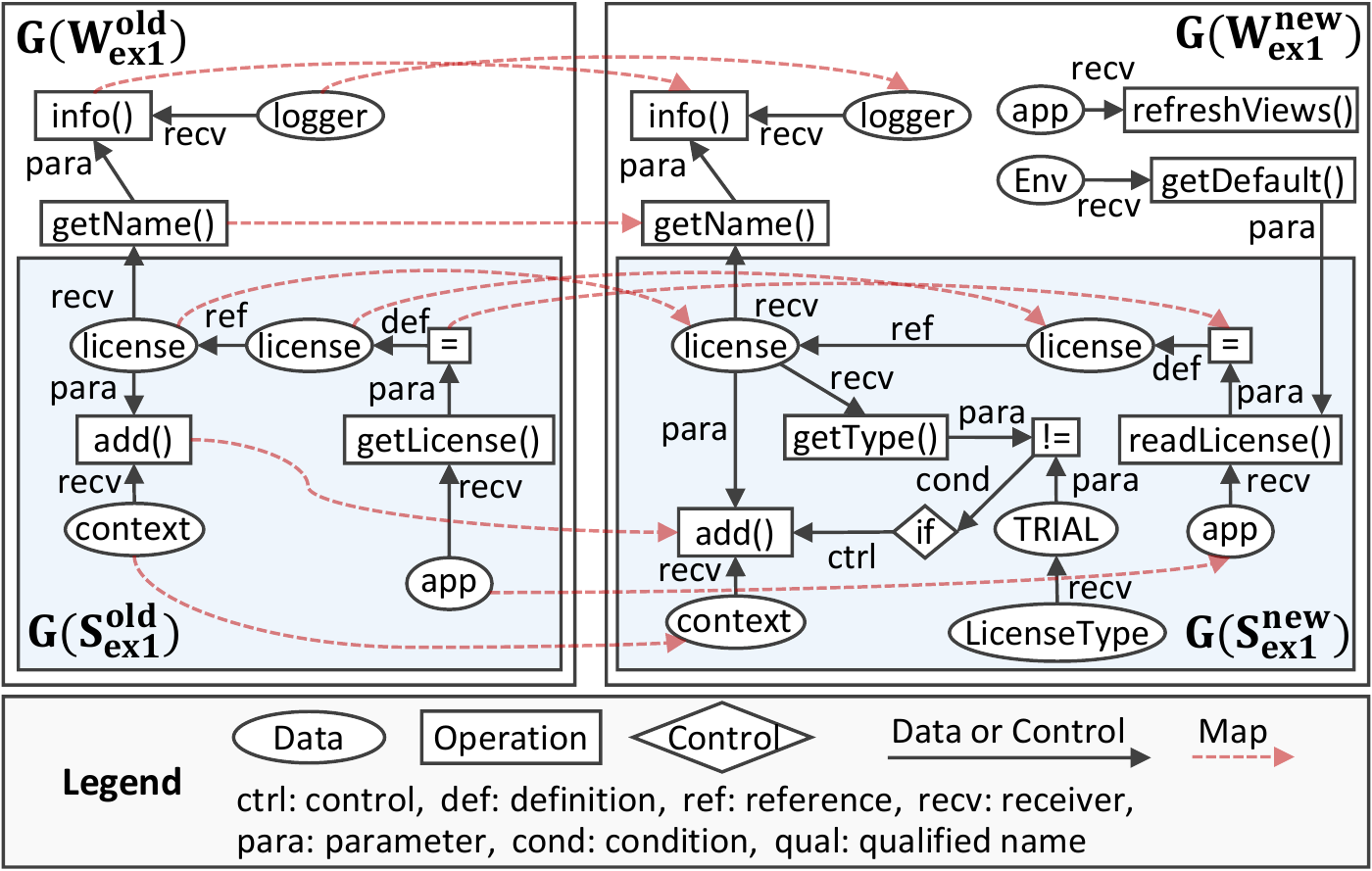}
  \caption{Change graph CG1 that represents code-change example 1 in \Cref{fig:code-delta1}.}
  \label{fig:change-graph1}
\end{figure}

The code change in \Cref{fig:code-delta1} is represented as change graph CG1 in \Cref{fig:change-graph1}.
Map edges linking old and new nodes are set on the basis of AST node mappings calculated using a traditional AST differencing algorithm.
Note that by definition transitive closure is computed, resulting in a graph with transitive edges, but we omit those edges in \Cref{fig:change-graph1} for simplicity.

After representing the code changes in \Cref{fig:code-delta2} as change graph CG2 in the same manner,
 we find SEPs in CG1 and CG2 by frequent subgraph mining.
The blue rectangle part in \Cref{fig:change-graph1} is common between CG1 and CG2; thus, it becomes an SEP
 (whose support in frequent subgraph mining is 2).

We use the following notations:
G(X) denotes an fgPDG of a code fragment X,
W stands for the whole code of a method,
and S stands for the code snippet corresponding to an SEP.
The superscripts `old' and `new' represent before and after a code change, and
the subscript `ex1' shows the method name.

\subsection{Difficulties of Program Repair with fgPDG-Based SEPs}
\label{sec:sep-application-difficulty}

To repair the overlooked (deprecated) code in the client, %
 we need to apply the same code changes as in the SEP to the deprecated location,
 i.e., replacing \textit{getLicense} with \textit{readLicense} and surrounding \textit{ctx.add} with an \textit{if} guard regarding license types.
However, straightforward ideas for achieving this code transformation result in several difficulties, as follows.%

\subsubsection{Use of Existing Transformation Techniques}
\label{sec:use-of-existing-techniques}
One possible idea for applying the SEP to the client is to leverage existing transformation techniques for syntax-based SEPs;
 however, they do not work well for the motivating example.
They first compute generalized AST edit scripts between old and new code in an SEP (i.e., a set of \code{W}{old}{} and \code{W}{new}{}).
The edit script is then applied to the client code.
The problem is that program structures of the pattern instances (\code{W}{}{ex1}, \code{W}{}{ex2}) and client (\code{W}{}{c}) mutually diverge;
 thus, it is quite difficult (or impossible) to compute such generalized AST edit scripts and apply them to the client code.

For example, a state-of-the-art transformation technique ARES\tcite{ares:fse2017} fails to infer an edit pattern from the change examples in \Cref{fig:code-delta1,fig:code-delta2},
 because there are many structural inconsistencies between \code{W}{}{ex1} and \code{W}{}{ex2}.
Another state-of-the-art technique GenPat\tcite{genpat:ase2019} fails to apply inferred transformation to \code{W}{}{c}.
It succeeds in replacing \textit{getLicense} with \textit{readLicense} but inserts an \textit{if} guard regarding license types to a wrong position.
It also inserts \textit{app.refreshViews()} that is not a part of the systematic edit to the client.
The existing techniques expect many structural similarities among pattern instances and client code, but such similarities are not guaranteed in PDG-based SEPs;
 there are many structural inconsistencies among the code structures of the motivating examples.

\subsubsection{PDG Transformation and Code Reconstruction}
\label{sec:difficulty-pdg-trans-code-recons}
Another idea is to first transform the client PDG such that it contains the same structure as the new one of the SEP (\fgpdg{S}{new}{ex1}),
 then reconstruct program code from the transformed PDG.

\Cref{fig:client-fgpdg} shows the fgPDG representation of the client \code{W}{}{c}.
The PDG transformation approach first replaces the subgraph \fgpdg{S}{}{c}, which is matched against \fgpdg{S}{old}{ex1}, with \fgpdg{S}{new}{ex1}.
The approach then reconstructs client code from the transformed PDG; however, this is infeasible due to the following reasons.
\begin{enumerate}
  \item An fgPDG does not contain all types of program elements (e.g., there are no PDG nodes corresponding to blocks and expression-statements);
        thus, it is impossible to recover program code from a transformed PDG.\label{difficulty-1}
  \item It is undecidable how to connect PDG nodes inside G(S) to ones outside G(S).
        For example, the replacement of \fgpdg{S}{}{c} with \fgpdg{S}{new}{ex1} results in two control edges to the \textit{add} node from the two \textit{if} nodes (originating from \fgpdg{W}{}{c} and \fgpdg{S}{new}{ex1}).
        It is impossible to determine which of the \textit{if} statements is inside the other in the source code.
        It is also a problem that there are no PDG nodes for the arguments of \textit{readLicense}.\label{difficulty-2}
\end{enumerate}

Sirius that incorporates our transformation algorithm for PDG-based SEPs avoids the difficulties mentioned above.
We elaborate on the details of Sirius in the next section.

\section{Sirius: Repair Pipeline with PDG-Based SEPs}
\label{sec:proposed-technique}

\begin{figure}[tb]
  \centering
  \includegraphics[width=.85\columnwidth]{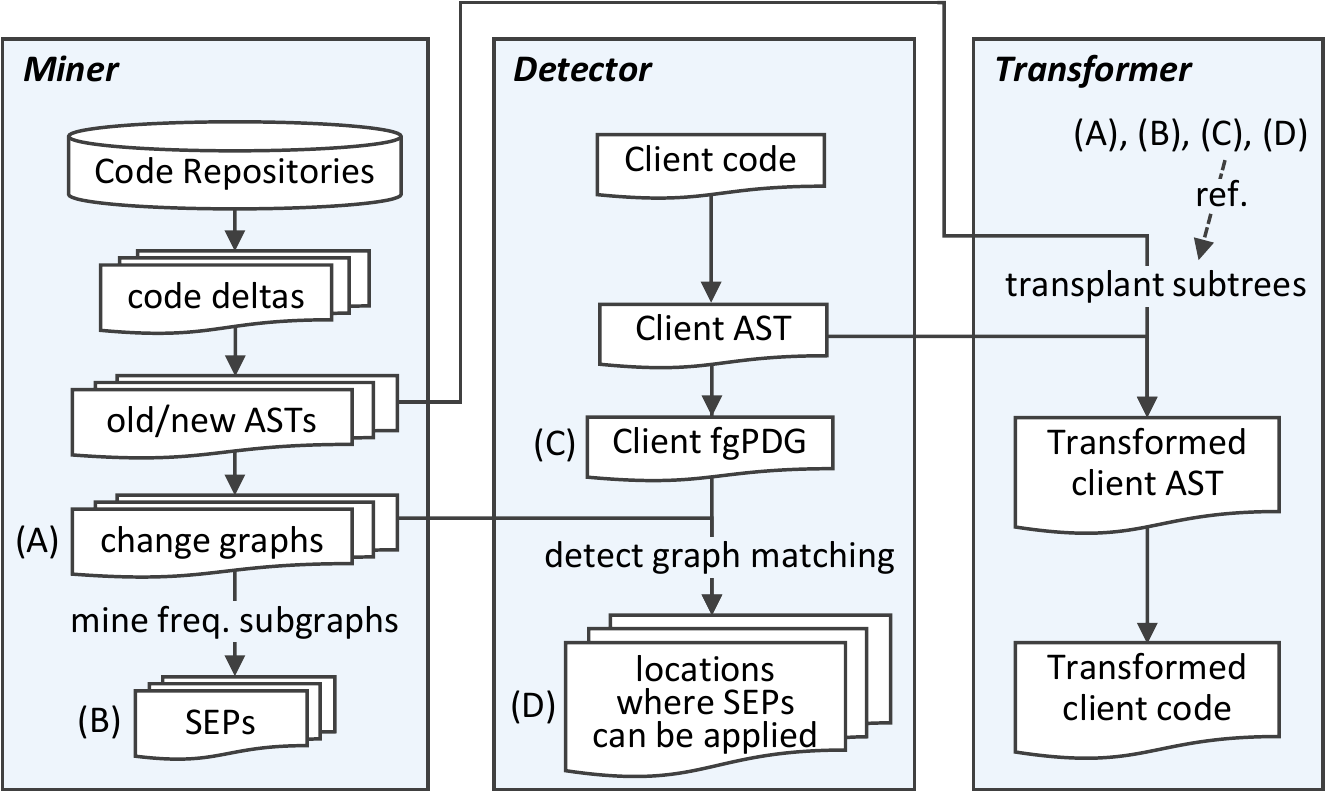}
  \caption{Overview of our repair pipeline \textit{Sirius}.}
  \label{fig:technique-overview}
\end{figure}

We present an automated program repair pipeline \textit{Sirius}.
\Cref{fig:technique-overview} shows an overview of Sirius.
There are three main components.
(1) \textit{Miner} mines fgPDG-based SEPs from development histories.
(2) \textit{Detector} detects locations where SEPs can be applied in the given client code.
(3) \textit{Transformer} applies the same code changes as in SEPs to the detected locations.

In the following, we denote the old version of a pattern instance code where an SEP is extracted as \code{W}{old}{p} and \code{S}{old}{p}.
In our motivating example, \code{W}{old}{p} and \code{S}{old}{p} correspond to \code{W}{old}{ex1} and \code{S}{old}{ex1}, respectively.
The same notations are used for the new version of a pattern instance code (i.e., \code{W}{new}{p} and \code{S}{new}{p}).

\subsection{Miner}

The miner extracts change graphs from development histories, then obtains PDG-based SEPs by frequent subgraph mining, which is the same manner as CPatMiner\tcite{cpatminer:icse2019}.
We slightly modify the representation (abstraction level) of fgPDGs used in CPatMiner to reduce false positives.

To obtain SEPs via frequent subgraph mining, we have to abstract (anonymize) some identifiers of program elements
 because local identifier names are generally not consistent among change graphs.
In our motivating example (\Cref{fig:code-delta1,fig:code-delta2}), variable names of \textit{License} instances differ from one another: \textit{license} and \textit{li}.
In Sirius (also in CPatMiner), the miner abstracts some types of node labels in fgPDGs.

\begin{figure}[tb]
  \centering
  \includegraphics[width=\columnwidth]{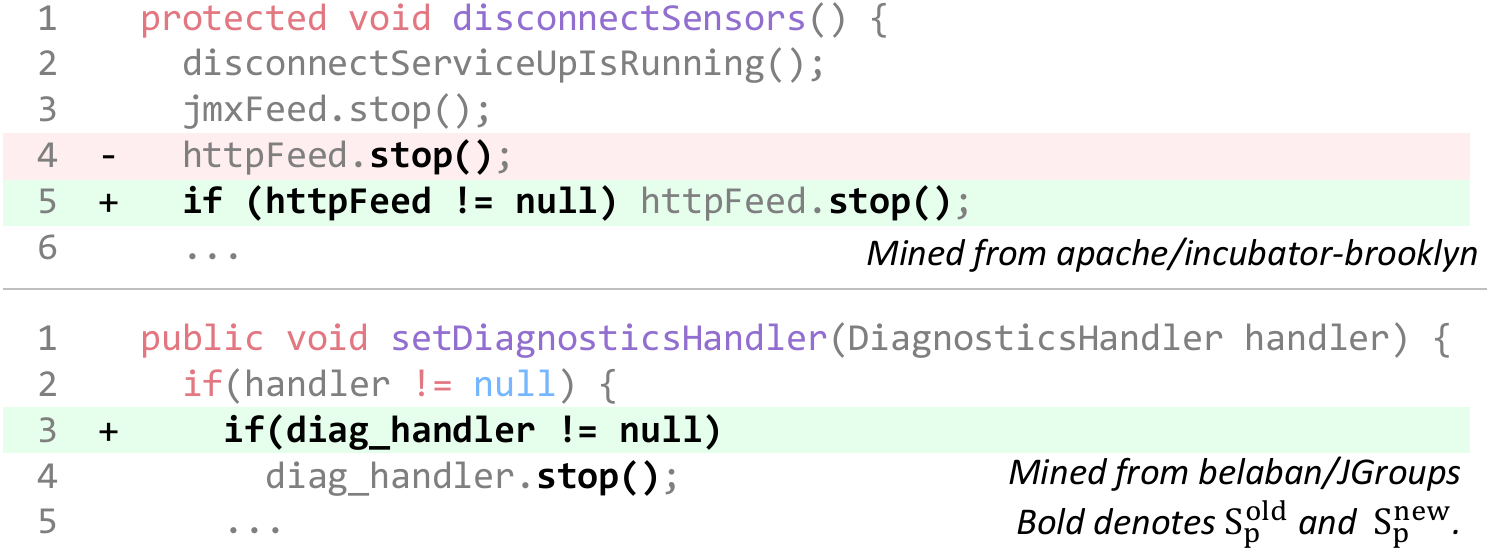}
  \caption{Example of a false positive reported by CPatMiner.}
  \label{fig:mined-sep2}
\end{figure}

\Cref{tab:comparison-node-label-abstraction} shows how to abstract node labels in fgPDGs.
CPatMiner regards all variable names as `*' (wildcard), literals as their data type names (number, null, etc.), and method invocations as their method names.
This higher level of abstraction leads to false positives in the mining results.
\Cref{fig:mined-sep2} shows an example of a false SEP reported by CPatMiner.
In \Cref{fig:mined-sep2}, invoked methods have a commonly used name \textit{stop}: therefore, the change deltas using methods from different classes are incorrectly regarded as the same systematic edit.

\begin{table}
  \centering
  \caption{Comparison of how to abstract node labels in fgPDGs.}
  \label{tab:comparison-node-label-abstraction}
  \begin{tabular}{llll}
    \hline
    Tool  & Variable & Literal & Method invocation \\
    \hline
    Sirius & type name & type name & receiver type + method name \\
    CPatMiner & * (wildcard) & type name & method name \\
    \hline
  \end{tabular}
\end{table}

To avoid this issue, Sirius leverages fgPDGs with a lower level of abstraction than CPatMiner.
Sirius abstracts variable and literal identifiers to their type names, and method invocations to their receiver types followed by the method names.
In our motivating example, the variables \textit{license} and \textit{li} are labeled \textit{License}, and the method invocations \textit{context.add} and \textit{ctx.add} are labeled \textit{Context\#add} in fgPDGs.

\subsection{Detector}

Given a client code (a repair target), the detector finds locations where SEPs can be applied.
To this end, it first converts the client code to an fgPDG (\fgpdg{W}{}{c}), then detects subgraphs in \fgpdg{W}{}{c} that are isomorphic to old graphs of SEPs (\fgpdg{S}{old}{p}).
Code elements corresponding to the detected subgraphs will be transformed on the basis of the SEPs by the transformer.
In our motivating example, \fgpdg{S}{}{c} in \Cref{fig:client-fgpdg} is detected as the location to transform by applying the SEP because it is isomorphic to the old graph of the SEP (\fgpdg{S}{old}{ex1}).

Some SEPs can be applied repeatedly to the same location.
For example, if we have an SEP $p=$ ``\textit{add(v)} $\rightarrow$ \textit{if (v != null) add(v)}'' and a client code $c=$ ``\textit{add(v)},''
 we can apply $p$ twice to $c$, resulting in ``\textit{if (v != null) if (v != null) add(v)}.''
To avoid such repetitive SEP application, the detector matches client graphs against not only old graphs of SEPs (\fgpdg{S}{old}{p}) but also new ones (\fgpdg{S}{new}{p}).
If a client graph matches against new graphs of SEPs, Sirius does not apply the SEPs to the client.
In the above example, once $p$ is applied to $c$, the transformed $c$ matches the new graph of $p$; we do not further apply $p$ to $c$.

\subsection{Transformer}
\label{sec:transformer}

\subsubsection{Overview}

The transformer applies the same change as in SEPs to code snippets corresponding to client subgraphs detected by the detector component.
In the following, A(X) denotes an AST corresponding to a code snippet X.

Since we cannot recover code or ASTs from fgPDGs (\Cref{sec:difficulty-pdg-trans-code-recons}-\ref{difficulty-1}),
 instead of directly transforming fgPDGs,
 our transformer identifies a set of transplantable AST nodes for each fgPDG node and transplants them from \astree{W}{new}{p} to \astree{W}{}{c}.
If we simply transplant each AST node corresponding to each fgPDG node individually to the client graph,
  we cannot determine how to connect transplanted nodes to the surrounding AST nodes of recipient locations;
  this is the same type of issue described in \Cref{sec:difficulty-pdg-trans-code-recons}-\ref{difficulty-2}.
To address this issue, we introduce \textit{Minimum Transplantable induced Subtree (MTS)} and transplant AST nodes in units of MTSs.

\subsubsection{Minimum Transplantable induced Subtree (MTS)}
\label{sec:mts}

In transplanting AST nodes from \astree{W}{new}{p} to \astree{W}{}{c},
 an MTS represents the smallest induced subtree such that we can determine the connections from the boundary (root and leaf) nodes of the MTS to AST nodes in \astree{W}{}{c}.
We call an AST node in \astree{W}{new}{p} \textit{nc-mapped} if there exists a corresponding AST node in \astree{W}{}{c}.
MTS is defined as follows.
\begin{df}
  Given an fgPDG node g in \fgpdg{S}{new}{p}, the MTS of g is denoted as MTS(g).
  MTS(g) is the smallest induced subtree in \astree{W}{new}{p} that satisfies all the following conditions.
  \begin{enumerate}
    \item MTS(g) contains an AST node in \astree{W}{new}{p} that corresponds to g;
    \item The root node of MTS(g) is nc-mapped or a statement node (e.g., expression-statement, if-statement, etc.);
    \item Each leaf of MTS(g) is nc-mapped or has no child.
  \end{enumerate}
\end{df}

In the following, we first elaborate on the mapping algorithm between \astree{W}{new}{p} and \astree{W}{}{c} that are used for checking whether each AST node is nc-mapped.
We then describe how to apply an SEP to the client via MTS-based transplantation.

\subsubsection{Mappings between \astree{W}{new}{p} and \astree{W}{}{c}}
\label{sec:pattern-new-client-mapping}

To compute MTSs, we need to know which AST node is nc-mapped, i.e.,
  the AST node mappings between \astree{W}{new}{p} and \astree{W}{}{c}.
 
Since \astree{W}{old}{p} and \astree{W}{new}{p} are the same method before and after a code change,
 most parts of the method are structurally consistent except for modified locations.
Thus, traditional AST differencing algorithms (e.g., GumTree\tcite{gumtree:ase2014}) work well for computing their AST node mappings.

On the other hand, \astree{W}{new}{p} and \astree{W}{}{c} generally originate from different methods.
There can be many structural inconsistencies in many parts of the methods.
Therefore, even if we apply traditional AST differencing algorithms, we may not be able to correctly obtain their AST node mappings.

The detector component guarantees that \fgpdg{S}{}{c} (i.e., client locations where SEPs can be applied) is isomorphic to \fgpdg{S}{old}{p}.
This means that AST substructures related to fgPDG nodes in \fgpdg{S}{}{c} and \fgpdg{S}{old}{p} should be locally consistent,
 although, globally, \code{W}{}{c} and \code{W}{old}{p} have many structural inconsistencies.
Leveraging this property, we compute AST mappings from \astree{W}{new}{p} to \astree{W}{}{c} through \fgpdg{W}{old}{p} and \fgpdg{W}{}{c}.

\Cref{alg:map-pattern-new-to-client} shows how to map AST nodes from \astree{W}{new}{p} to \astree{W}{}{c}.
We assume that mappings between \astree{W}{new}{p} and \astree{W}{old}{p} (denoted as \mapping{A}{new-old}) have been computed in advance with traditional AST differencing algorithms such as GumTree.
Besides, mappings of fgPDG nodes inside an SEP (i.e., inside \fgpdg{S}{old}{p} and \fgpdg{S}{}{c}) are given from the detector component and denoted as \mapping{G(S)}{old-cli}.

The algorithm transitively traverses mappings from the new version of a pattern instance (\code{W}{new}{p}),
 through its old version (\code{W}{old}{p}), to the client (\code{W}{}{c}).
 During the traversal, the mappings between \code{W}{old}{p} and \code{W}{}{c} that are originally given by the detector are augmented on the basis of path similarities in fgPDGs.
 As the result of the algorithm, all transitive mappings from \astree{W}{new}{p} to \astree{W}{}{c} are collected in its output (\mapping{A}{new-cli}).

\begin{algorithm}[tb]
  \caption{Map nodes in \astree{W}{new}{p} to nodes in \astree{W}{}{c}.}\label{alg:map-pattern-new-to-client}
  \begin{ls_algorithmic}
    \Require \mapping{A}{new-old}: AST node mappings b/w \astree{W}{new}{p} and \astree{W}{old}{p};
    \Statex \hspace{1.1em} \mapping{G(S)}{old-cli}: fgPDG node mappings b/w \fgpdg{S}{old}{p} and \fgpdg{S}{}{c}.
    \Ensure \mapping{A}{new-cli}: AST node mappings b/w \astree{W}{new}{p} and \astree{W}{}{c}.
    \State \tr{allOldG} $\gets \{ $ g $ \mid \tr{n} \in$ \astree{W}{new}{p} s.t. getM(n, \mapping{A}{new-old}) $\neq$ null,
    \Statex \hspace{6.8em} g $\in$ getGNodes(getM(n, \mapping{A}{new-old}))\}
    \State \mapping{G}{old-cli} $\gets$ detectOldToClientGMap(\tr{allOldG}, \mapping{G(S)}{old-cli})
    \State \mapping{A}{new-cli} $\gets$ detectNewToClientMap(\astree{W}{new}{p}, \mapping{A}{new-old}, \mapping{G}{old-cli})
    \State \Return \mapping{A}{new-cli}
    \Statex
    \Function{detectOldToClientGMap}{allOldG, \mapping{G(S)}{old-cli}}
      \Statex $\triangleright$ \textbf{In:} allOldG: fgPDG nodes in \fgpdg{W}{old}{p};
      \Statex \hspace{2.2em} \mapping{G(S)}{old-cli}: fgPDG node mappings b/w \fgpdg{S}{old}{p} and \fgpdg{S}{}{c}.
      \Statex $\triangleright$ \textbf{Out:} \mapping{G}{old-cli}: fgPDG node mappings b/w \fgpdg{W}{old}{p} and \fgpdg{W}{}{c}.
    \State oldG${}^{\tr{W}\setminus\tr{S}}$ $\gets$ \{g $\mid$ g $\in$ allOldG $\land$ g $\not \in$ \fgpdg{S}{old}{p} \}
    \State clientG${}^{\tr{W}\setminus\tr{S}}$ $\gets$ \{g $\mid$ g $\in$ \fgpdg{W}{}{c} $\land$ g $\not \in$ \fgpdg{S}{}{c} \}
    \State C\mapping{G(W$\setminus$S)}{old-cli} $\gets$ calcPathBasedGMap(oldG${}^{\tr{W}\setminus\tr{S}}$, clientG${}^{\tr{W}\setminus\tr{S}}$)
    \State \mapping{G(W$\setminus$S)}{old-cli} $\gets$ solveMaxCardinalityBipartiteMatching(C\mapping{G(W$\setminus$S)}{old-cli})
    \State \Return \mapping{G(S)}{old-cli} $\cup$ \mapping{G(W$\setminus$S)}{old-cli}
    \EndFunction
  \end{ls_algorithmic}
\end{algorithm}

In \Cref{alg:map-pattern-new-to-client},
 we first identify fgPDG nodes in \fgpdg{W}{old}{p} that correspond to each AST node in \astree{W}{new}{p} via \mapping{A}{new-old} (l.1).
getM($n, M$) returns the (AST or PDG) node mapped from the given node $n$ via the mapping $M$.
 If $n$ has no mapped node, it returns null.
getGNodes($a$) takes an AST node $a$ in A(X) and returns the corresponding fgPDG nodes in G(X).

We then obtain mappings between \fgpdg{W}{old}{p} and \fgpdg{W}{}{c} by detectOldToClientGMap (l.2).
 Since the mappings between fgPDG nodes inside an SEP have already been identified by the detector,
 it needs to compute only the mappings of fgPDG nodes outside the SEP (i.e., outside \fgpdg{S}{old}{p} and \fgpdg{S}{}{c}).
To obtain the mappings, we compute path-based similarities for every pair of fgPDG nodes outside the SEP (ll.6--9).

  At line 8, we encode every fgPDG node $g$ outside the SEP to a path set $\tr{P}(g) = \{ p \mid p \gets \langle \tr{L}(g), \tr{L}(e), \tr{L}(\textit{dest}) \rangle \}$.
  $\tr{P}(g)$ enumerates every path $p$ from the node $g$ outside the SEP to a \textit{dest} node inside the SEP via the edge $e$.
  L returns the label of the given node or edge.
  We calculate set similarities (e.g., Dice's coefficient) for every pair of P($g_1$ $\in$ oldG${}^{\tr{W}\setminus\tr{S}}$) and P($g_2$ $\in$ clientG${}^{\tr{W}\setminus\tr{S}}$),
   thereby detecting candidate node mappings between oldG${}^{\tr{W}\setminus\tr{S}}$ and clientG${}^{\tr{W}\setminus\tr{S}}$.
  Each node $g_1$ is mapped to the node(s) $g_2$ having the highest similarity score.
  Since $g_1$ can be mapped to more than one fgPDG node, we determine the final one-to-one mappings by maximum-cardinality matching in a bipartite graph of the two node-sets, oldG${}^{\tr{W}\setminus\tr{S}}$ and clientG${}^{\tr{W}\setminus\tr{S}}$ (l.9).
  In our motivating example, for example, \textit{getName} in \fgpdg{S}{old}{ex1} is encoded as $\{ \langle \textit{getName}, \textit{recv}, \textit{license} \rangle \}$.
  The client node \textit{getName} in \fgpdg{S}{}{c} has the same encoded value.
  Thus, the two nodes are mutually mapped.

Finally, we identify the AST node in \astree{W}{}{c} that corresponds to each AST node in \astree{W}{new}{p}
 by transitively traversing the two mappings, \mapping{A}{new-old} and \mapping{G}{old-cli} (l.3).
Note that the correspondence between fgPDG nodes in G(X) and AST nodes in A(X) are easily identified because each fgPDG is constructed from the corresponding AST.
For an AST node in \astree{W}{new}{p}, if the traversal reaches multiple AST nodes in \astree{W}{}{c}, we select the mode value (appearing most often).

\subsubsection{MTS-Based SEP Application}
\label{sec:sep-application-based-on-mts}

\Cref{alg:transplant-mts-to-client} shows how to apply an SEP to the client AST on the basis of MTSs.
It deletes the code element (AST node) corresponding to each old node in the pattern graph (\fgpdg{S}{old}{p}) from the client.
Then, it inserts code elements corresponding to each new node in the pattern graph (\code{S}{new}{p}) to the client in units of MTSs.
Consequently, the algorithm ensures that no code elements in \code{S}{old}{p} and all the ones in \code{S}{new}{p} are present in the resulting client code.

\begin{algorithm}[tb]
  \caption{Apply an SEP to \astree{W}{}{c}.}\label{alg:transplant-mts-to-client}
  \begin{ls_algorithmic}
    \Require \mapping{A}{new-cli}: AST node mappings b/w \astree{W}{new}{p} and \astree{W}{}{c}.
    \Ensure \astree{W}{new}{c}: the client AST transformed by applying an SEP.
    \State Delete every AST node in \astree{W}{}{c} that corresponds to g $\in$ \fgpdg{S}{}{c}.
    \State mtsSet $\gets$ \{ MTS($\tr{g}_i$, \mapping{A}{new-cli}) $\mid$ $\tr{g}_i$ $\in$ \fgpdg{S}{new}{p}
    \Statex \hspace{3.5em} $\land$ $\forall j (\neq i)$, MTS($\tr{g}_i$, \mapping{A}{new-cli}) $\not\subseteq$  MTS($\tr{g}_j$, \mapping{A}{new-cli}) \}
    \ForEach{mts $\in$ mtsSet}
    \State clientR $\gets$ getM(mts.rootNode, \mapping{A}{new-cli})
    \If{clientR $\neq$ null}
    \State connectASTNodes(clientR.parent, \{mts.rootNode\})
    \Else
    \State Let c $\in$ mts be a nc-mapped node;\ \ m $\gets$ getM(c, \mapping{A}{new-cli});
    \State connectASTNodeToAncestor(m, mts.rootNode)
    \EndIf
    \ForEach{l $\in$ mts.leafNodes}
    \State clientL $\gets$ getM(l, \mapping{A}{new-cli})
    \If{clientL $\neq$ null}
    \State connectASTNodes(l, clientL.children)
    \EndIf
    \EndFor
    \EndFor
    \Statex
    \Function{MTS}{g, \mapping{A}{new-cli}}
    \State a $\gets$ the AST node corresponding to g;$\ $ $\tr{S}_\tr{MTS} \gets \{ \tr{a} \}$;
    \While{(a is not nc-mapped $\land$ a is not a statement)}
    \State a $\gets$ a.parent;$\ $ $\tr{S}_\tr{MTS} \gets \tr{S}_\tr{MTS} \cup\ \{ \tr{a} \}$;
    \EndWhile
    \State bfsQueue $\gets \{ \tr{c} \mid \tr{c} \in \tr{a.children} \}$
    \While{(bfsQueue $\neq \emptyset$)}
    \State c $\gets$ bfsQueue.poll();$\ $ $\tr{S}_\tr{MTS} \gets \tr{S}_\tr{MTS} \cup\ \{ \tr{c} \}$;
    \If{c is not nc-mapped}
    \State bfsQueue $\gets$ bfsQueue $\cup\ \{ \tr{c}' \mid \tr{c}' \in \tr{c.children} \}$
    \EndIf
    \EndWhile
    \State \Return $\tr{S}_\tr{MTS}$
    \EndFunction
  \end{ls_algorithmic}
\end{algorithm}

First, we delete AST nodes in \astree{W}{}{c} that correspond to fgPDG nodes in \fgpdg{S}{}{c} (l.1).
After that, we compute an MTS for each node in \fgpdg{S}{new}{p}, resulting in a set of MTSs to transplant (l.2).
The MTS function returns the minimum induced subtree that satisfies the conditions described in \Cref{sec:mts} (ll.14--23).
An MTS contained in another MTS is ignored (l.2).

We then connect the boundary nodes of each MTS to the parent or leaves of the corresponding AST nodes in \astree{W}{}{c} (ll.3--13).
connectASTNodes($p$, $N$) connects every AST node n $\in$ N as the children of $p$.
The index of $n$ in the child list of $p$ is the same as that in the AST tree having $n$.
Note that $n$ can be an expression; in that case, we wrap $n$ in a newly created expression-statement node to connect as a child of $p$.

If the root node of an MTS (mts.rootNode) is not nc-mapped (i.e., a statement node),
 we first select a node $m$ in \astree{W}{}{c} mapped from an nc-mapped node in the MTS (l.8).
 We then find the $m$'s nearest ancestor that can have statements as its children,
 and connect mts.rootNode to that node (l.9).
Note that if a statement that is not nc-mapped cannot become the root of an MTS, the size of an MTS enlarges, increasing the risk of excessive transplantation from \astree{W}{new}{p} to \astree{W}{}{c}.
Thus, we allow unmapped statements to be MTS root nodes.

We show some concrete examples of MTS transplantation.%

\begin{figure}[tb]
  \centering
  \includegraphics[width=\columnwidth]{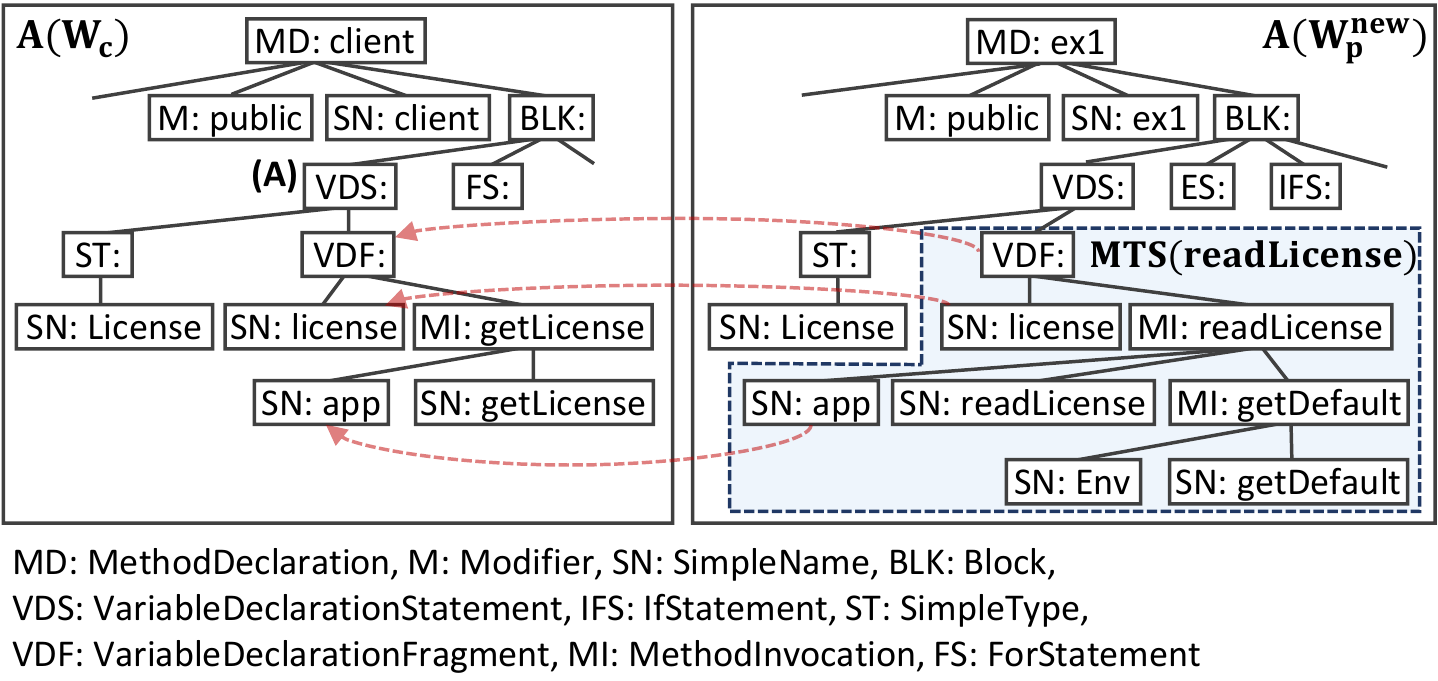}
  \caption{Excerpt of AST related to ll.3--4 in \Cref{fig:code-delta1}.}
  \label{fig:mts-transplantation1}
\end{figure}

As an example, we apply the change of ll.3--4 in \Cref{fig:code-delta1} to the client: changing \textit{getLicense} to \textit{readLicense}.
\Cref{fig:mts-transplantation1} shows the relevant parts of \astree{W}{}{c} and \astree{W}{new}{p}.
First, we delete AST nodes in \astree{W}{}{c} that correspond to the fgPDG nodes in \fgpdg{S}{}{c}.
For example, `VDF:' and `MI: getLicense' nodes are deleted.
Then, computing the MTS of \textit{readLicense} in \fgpdg{S}{new}{p},
we connect the root of the MTS to the parent of its mapped node (i.e., (A) in \Cref{fig:mts-transplantation1}).
For the leaves of the MTS, since the `SN: app' node has no children, there is no need to set up any connections.
As a result, the `VDS' node ((A) in \Cref{fig:mts-transplantation1}) has the MTS as its child, resulting in the transformed client code to which the change of ll.3--4 in \Cref{fig:code-delta1} has been applied.

\begin{figure}[tb]
  \centering
  \includegraphics[width=.9\columnwidth]{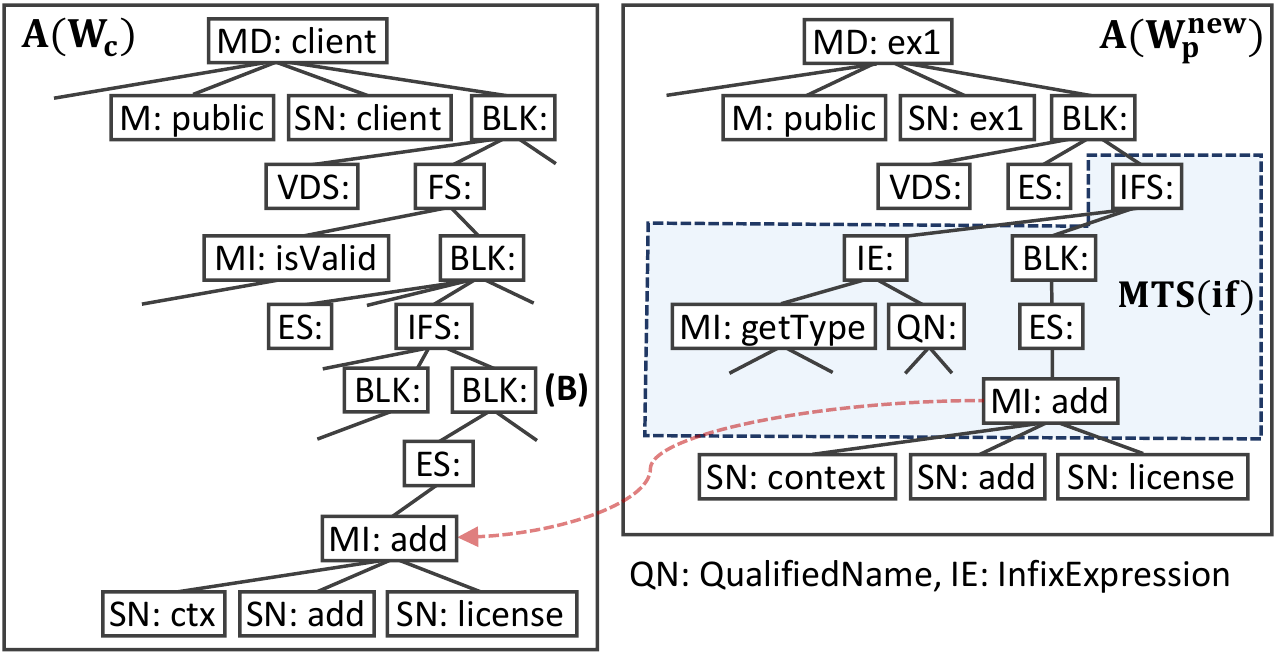}
  \caption{Excerpt of AST related to ll.6--8 in \Cref{fig:code-delta1}.}
  \label{fig:mts-transplantation2}
\end{figure}

As another example, we apply the change of ll.6--8 in \Cref{fig:code-delta1} to the client: adding an \textit{if} statement regarding license types.
\Cref{fig:mts-transplantation2} shows the relevant parts of \astree{W}{}{c} and \astree{W}{new}{p}.
Since the root $r$ of MTS(if) is not mapped,
 we select an nc-mapped node `MI: add' in the MTS,
 get its mapped node,
 find the nearest ancestor \textit{na} that can have statements as its child (i.e., $na$ is (B) in \Cref{fig:mts-transplantation2}),
 and replace the $na$'s child with $r$.
We then set up connections for the mapped leaf nodes in the MTS; 
 we replace the children of `MI: add' in the MTS(if) with those of `MI: add' in \astree{W}{}{c}.
As a result, \textit{add} in \code{W}{}{c} is replaced with \textit{add} surrounded by the \textit{if} guard regarding license type.

\subsubsection{Transplantation of Variables and Literals}
\label{sec:label-concretization}

Since \code{W}{new}{p} and \code{W}{}{c} are generally different methods, variable and literal names are inconsistent between the two methods.
In MTS-based transplantation, we need to determine which names are appropriate to use.
We also have to sometimes avoid conflicts of names between the two methods.

As a general policy, we try to use (preserve) names used in \code{W}{}{c}.
Specifically, we determine the name of each variable or literal X in MTSs as follows in the following order of priority.
\begin{enumerate}
  \item If there are fgPDG nodes in \fgpdg{W}{old}{p} that have the same name as X, we try to find the corresponding node CG in \fgpdg{W}{}{c} via \mapping{G}{old-cli}.
        If found, we use the name of CG for the name of X.
  \item If variable declaration fragments for X are transplanted from \code{W}{new}{p} to \code{W}{}{c}, we use (i.e., does not change) the name of X after transplantation.\label{preserve-name1}
  \item If X corresponds to an AST node that is nc-mapped to an AST node CA in \astree{W}{}{c}, we use the name of CA for the name of X.
  \item Otherwise, we use (i.e., does not change) the name of X after transplantation.\label{preserve-name2}
\end{enumerate}

In \ref{preserve-name1} and \ref{preserve-name2}, if the name of X has already been used for a variable in \code{W}{}{c} before transplantation,
 we rename X with a fresh name to avoid a variable-name conflict.

\section{Evaluation Design}
\label{sec:evaluation-design}

\subsection{Research Questions}
\label{sec:research-question}

We evaluate Sirius in terms of effectiveness and efficiency.

\begin{mdframed}[style=simple-frame]
\textbf{RQ1:}
How effective is Sirius in repairing code that requires systematic edits represented by fgPDG-based SEPs?
\end{mdframed}

Our primary goal is to repair (overlooked) code locations that require systematic edits by leveraging fgPDG-based SEPs.
For RQ1, we evaluate the performance of Sirius in automated program repair for such code locations.

We conduct cross-validation to test whether Sirius can correctly repair code by applying real-world systematic edits.
We measure the overall repair performance as precision, recall, and F1-score through this cross-validation.
The details of the cross-validation setting are described in later sections.

Besides, we also investigate the performance against the graph sizes of patterns we exploit.
We consider that the more complex the change represented by a pattern is, the more difficult it is to apply the pattern.
The complexity of changes can be (indirectly) captured by the graph size (\#nodes) of each SEP;
 we need to edit many AST substructures when there are many nodes in a change graph.
Thus, there is a possibility that Sirius can achieve higher performance (esp., precision) by exploiting only small patterns with sizes less than or equal to a threshold $t$, which is practically valuable.
We measure the change of the performance when the size threshold $t$ varies.

We also sample failure cases from the cross-validation results to examine the causes of the failures.

\vspace{0.5em}
\begin{mdframed}[style=simple-frame]
\textbf{RQ2:}
How efficiently (fast) can Sirius produce program transformation results?
\end{mdframed}

In terms of practicality, time efficiency is also important as well as repair performance.
For this research question, we evaluate the time efficiency of Sirius
 by measuring the time spent in each phase: mining, detection, and transformation.

\subsection{Dataset}
\label{sec:evaluation-dataset}

We use the same dataset of OSS as for evaluating CPatMiner\tcite{cpatminer:icse2019}.
There are two corpora provided by the authors of CPatMiner: \ti{survey} and \ti{depth}.
The survey corpus was built for surveying developer responses,
 and consists of only fresh (the latest 50) commits of many projects to obtain good responses from developers about their recent changes.
The depth corpus was built for in-depth analysis of pattern characteristics,
 and has longer histories (the latest 1K commits with Java file changes) of fewer but high-quality projects.

We select the depth corpus as our dataset because the purpose of our experiment is not to obtain developer responses
 but to measure the repair performance and investigate the details of the results.
The depth corpus contains 88 real-world projects in various domains.
Every project has at least 10 committers and 100 GitHub stars.
The corpus contains a total of 1M Java files and 164M LOC.

\subsection{Baseline}
\label{sec:baseline}

We select ARES\tcite{ares:fse2017}, a state-of-the-art program transformation technique for syntax-based SEPs, as our baseline.
ARES takes change examples, infers transformation patterns, and applies the transformation to other code locations.
ARES achieved higher accuracy in recommending code changes and outperformed the previous state-of-the-art technique.

ARES was originally evaluated in two different settings\tcite{ares:fse2017}.
The first one uses two change examples for inferring more precise transformation patterns.
The other one takes all change examples to increase recall, which might result in overgeneralization (less accurate results).
Following that study, we also use ARES in the same two settings, which we call ARES-2 and ARES-all, respectively.

There is another recent publicly-available transformation technique, GenPat\tcite{genpat:ase2019},
 which learns program transformation from one change example while Sirius and ARES require several (many) change examples to learn.
The advantage of GenPat is that it can learn from rarely-conducted systematic edits.
However, learning from just one example tends to result in wrong and excessive transformation as described in \Cref{sec:use-of-existing-techniques}.
Direct comparison between GenPat and Sirius is not fair.
Thus, we do not select GenPat as our baseline.

\subsection{Procedure}
\label{sec:evaluation-procedure}

We conduct cross-validation to measure the repair performance.
First, we mine fgPDG-based SEPs from our dataset, resulting in a pattern collection \textit{PC}.
To obtain the \textit{PC}, we run the pattern miner with the minimum support of 3
 because, as described below, each pattern must have at least three pattern instances: more than two change examples as training data and one example as test data.
With \textit{PC}, we conduct leave-one-out cross-validation that simulates a systematic edit for each pattern instance,
 which reports the repair performance.

Let \ti{PC} be a set of mined fgPDG-based SEPs: $\ti{PC} = \{ p_1, p_2, ... \}$ where $p_i$ is an fgPDG-based SEP.
Each pattern $p_i$ has a set of pattern instances $\{ e^{p_i}_1, e^{p_i}_2, ... \}$.
In our motivating example, the mined SEP has two pattern instances: one originates from \code{W}{}{ex1} and the other originates from \code{W}{}{ex2}.

Let $C(e_i)$ be a function that takes a pattern instance $e_i$ and returns its \textit{change-id}, a pair of $\langle m, t \rangle$.
$m$ is the method where the change graph of $e_i$ is extracted.
$t$ is the commit time of the commit where $m$ was modified and $e_i$ is extracted.
With $C(e_i)$, for each pattern $p_i$, we can enumerate the change-ids of all pattern instances: $\textit{CS} = \{ C(e^{p_i}_1), C(e^{p_i}_2), ... \}$.

We conduct leave-one-\textit{change-id}-out cross-validation that takes into account the chronological order of commits.
Given test data (i.e., a change-id) $\langle m, t \rangle$, the training data are all change-ids $\{ \langle m_\textit{j1}, t_\textit{j1} \rangle, \langle m_\textit{j2}, t_\textit{j2} \rangle, ... \} \subset \textit{CS}$
 where $\forall k, t_\textit{jk} \le t$ (i.e., future commits are not used as training data).

One iteration in our cross-validation is processed as follows.
Let $C(e_\tr{test}^p) = \langle m, t \rangle$ be test data and $\ti{CS}_\tr{train}$ be training data
 where $p \in \ti{PC}$ is an fgPDG-based SEP.
We denote the code of $m$ before (resp.\ after) the commit at the time $t$ as $m_\tr{old}$ (resp.\ $m_\tr{new}$);
 $m_\tr{old}$ is the repair target and $m_\tr{new}$ contains the ground truth code of the repair.
Sirius attempts to mine an fgPDG-based SEP $p'$ from $\ti{CS}_\tr{train}$,
 detect code locations in $m_\tr{old}$ where $p'$ can be applied,
 and repair the detected location by applying $p'$.
Similarly, ARES attempts to infer a transformation pattern from $\ti{CS}_\tr{train}$ and apply it to $m_\tr{old}$ for repair.
We denote the repair result of Sirius or ARES as $m_\tr{repaired}$.

We judge that the repair result $m_\tr{repaired}$ is correct if all the following conditions are satisfied:
  (1) $m_\tr{repaired}$ contains all the code elements $\ti{CE}_\tr{new}$ in $m_\tr{new}$ such that $\ti{CE}_\tr{new}$ corresponds to the pattern's new graph (\fgpdg{S}{new}{p});
  (2) $m_\tr{repaired}$ does not contain any code elements $\ti{CE}_\tr{old}$ in $m_\tr{old}$ such that $\ti{CE}_\tr{old}$ corresponds to the pattern's old graph (\fgpdg{S}{old}{p});
  and (3) $m_\tr{repaired}$ contains all the code elements that are unchanged between $m_\tr{old}$ and $m_\tr{new}$.
Conditions (1) and (2) check successful insertion and deletion of pattern-related code, respectively.
Condition (3) ensures that there are no excessive modifications.

We check conditions (1) and (2) by testing whether the fgPDG of $m_\tr{repaired}$ contains \fgpdg{S}{new}{p} and does not contain \fgpdg{S}{old}{p}, respectively.
Also, we check condition (3) by AST/token-based comparison.
Note that if $m_\tr{repaired}$ is syntactically invalid, we cannot build its AST/fgPDG, regarding $m_\tr{repaired}$ as an incorrect result (repair failure).

The above evaluation logic relies on fgPDG-based comparison.
Although fgPDGs constitute the basis for Sirius's transformation algorithm, the fgPDG-based evaluation logic does not unfairly advantage Sirius.
First, $p'$ used for Sirius's transformation is different from $p$ used for evaluation.
Sirius (and ARES) does not know anything about $p$ during repair.
Besides, the transformation algorithm of Sirius transplants not fgPDG (sub)structures but partial ASTs (MTSs) to the client
 (e.g., Sirius does not transplant dependence edges).
Thus, our evaluation logic (fgPDG-based comparison) is independent of the transformation logic of Sirius,
 meaning that it does not give any advantages to Sirius.
Note that we cannot check conditions (1) and (2) by AST-based comparison
 that compares the code difference between $m_\tr{old}$ and $m_\tr{new}$ with that between $m_\tr{old}$ and $m_\tr{repaired}$,
 because the difference between $m_\tr{old}$ and $m_\tr{new}$ contains pattern-unrelated changes (e.g., the yellow-colored lines in \Cref{fig:code-delta1,fig:code-delta2}).
Checking (1) and (2) intrinsically requires fgPDG-based comparison.

\section{Evaluation Results}
\label{sec:evaluation-results}

\subsection{RQ1: Effectiveness}

First, we ran the pattern miner on the entire set of projects in the dataset.
The miner analyzed 288,497 change graphs, then produced 6,060 fgPDG-based SEPs.
We then conducted cross-validation using the mined patterns to measure the repair performance of Sirius and the baselines.
This involved a total of 31,273 repair trials for each technique.

\subsubsection{Overall Performance}
\Cref{tab:overall-repair-performance} shows the overall results of the cross-validation.
Sirius greatly outperforms the baselines for precision, recall, and F1-score.
The total numbers of generated transformed code are shown in parentheses in \Cref{tab:overall-repair-performance}.
Note that, for each repair trial in the cross-validation, the number of transformed code generated by ARES can be more than one while that of Sirius is at most one.

\begin{table}[tb]
  \centering
  \caption{Overall repair performance.}
  \label{tab:overall-repair-performance}
  \begin{tabular}{llll}
    \hline
    Tool & Precision & Recall & F1-score \\
    \hline
    Sirius & \textbf{\Sprecision{}} (17,683/24,889) & \textbf{\Srecall{}} (17,683/31,273) & \textbf{\Sfscore{}} \\
    ARES-2 & 0.470 (4,600/9,778) & 0.141 (4,395/31,273) & 0.216 \\
    ARES-all & 0.166 (4,335/26,168) & 0.123 (3,832/31,273) & 0.141 \\
    \hline
  \end{tabular}
\end{table}

ARES cannot well infer transformation patterns for many fgPDG-based SEPs because there are many structural inconsistencies among pattern instances and client code, as described in \Cref{sec:motivating-example}.
As the number of change examples for inferring transformation increases, such inconsistencies are more likely to occur, which is one of the main reasons that the performance of ARES-all is lower than that of ARES-2.

On the other hand, Sirius can handle such inconsistent cases thanks to MTS-based transplantation that requires less structural similarity.

Since ARES-2 outperforms ARES-all, we compare Sirius with ARES-2 in the following.

\subsubsection{Performance Against Graph-Size Threshold}
\label{sec:perf-against-graph-size-threshold}

\begin{figure}[tb]
  \begin{minipage}{.5\columnwidth}
  \centering
    \includegraphics[width=.95\columnwidth]{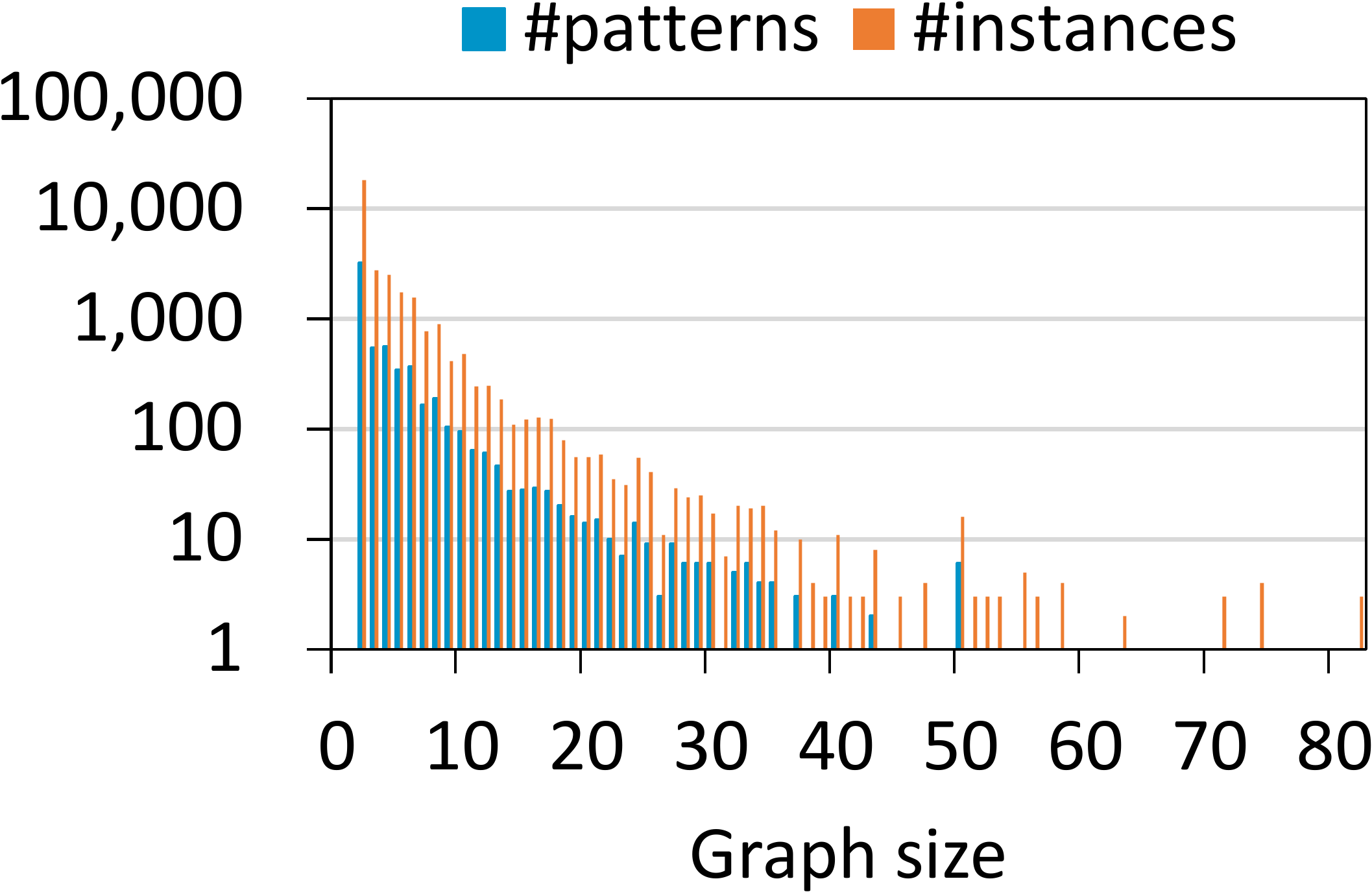}
  \end{minipage}
  \begin{minipage}{.48\columnwidth}
    \centering
    \includegraphics[width=.95\columnwidth]{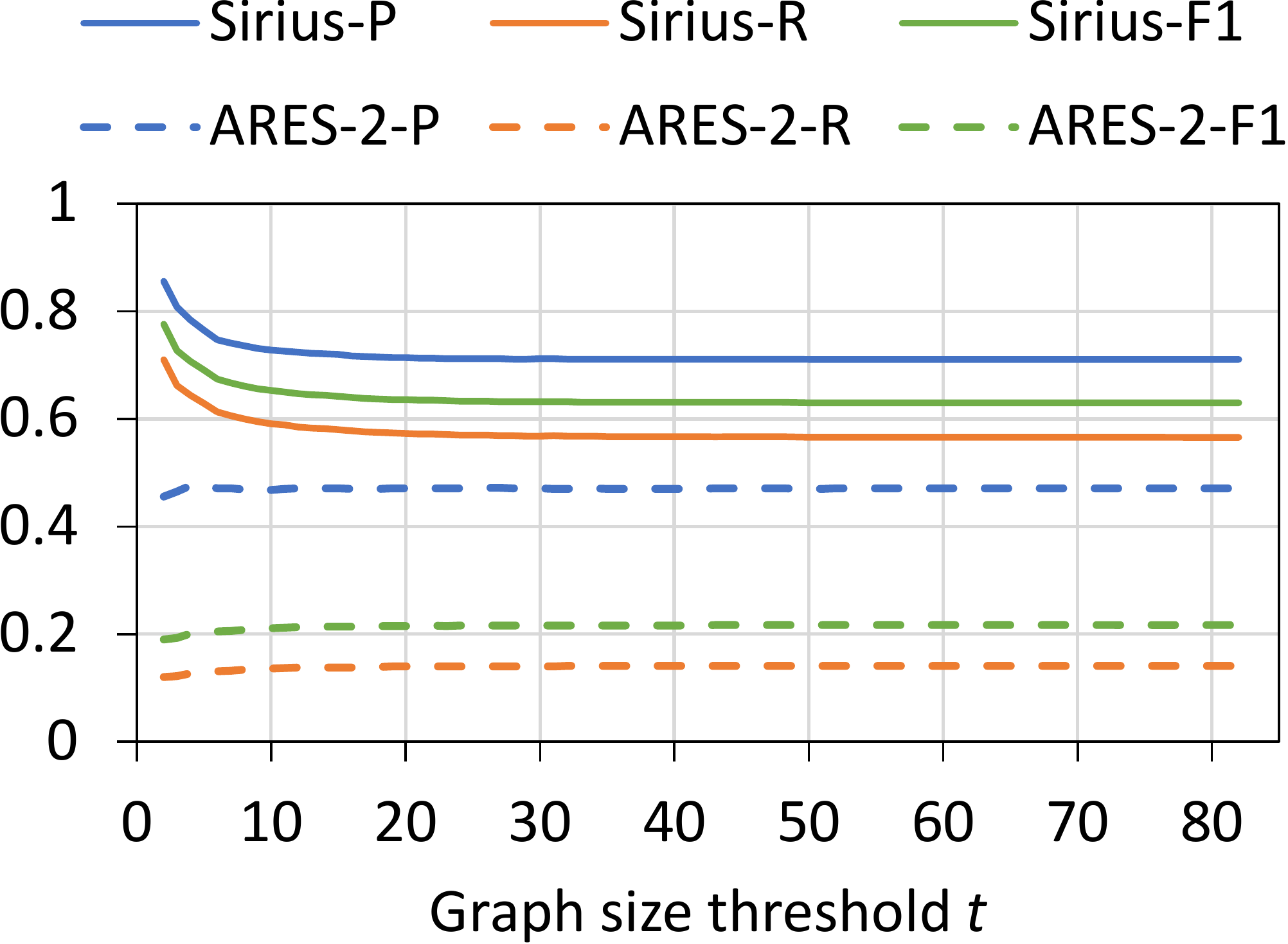}
  \end{minipage}
  \caption{Pattern distribution and repair performance against graph sizes of patterns.
  The left figure shows the numbers of patterns and their instances for each graph size.
  The right figure shows the repair performance for patterns with sizes of $t$ or less.
  P and R denote precision and recall, respectively.}
  \label{fig:performance-details}
\end{figure}

The left figure in \Cref{fig:performance-details} shows the distribution of the number of mined patterns and their instances.
The right figure in \Cref{fig:performance-details} shows the repair performance of each technique for patterns with sizes of $t$ or less.

Sirius achieves higher performance compared with ARES at any point of the threshold $t$.
Especially, Sirius has good precision ($\approx$ 0.8) and recall ($\approx$ 0.6) for small sizes of patterns.
Since patterns with a size of 10 or less account for around 90\% of the total, this is valuable for practical use.

The performance of Sirius decreases as the size threshold $t$ increases.
This is because the complexity of transplantation increases for larger pattern graphs where we need to edit much more AST nodes.
On the other hand, as for ARES, we cannot observe any trends in the performance change against the graph size threshold.
Since ARES does not use the pattern graphs, the complexity of transformations in ARES does not necessarily depend on graph size.

\begin{figure}[tb]
  \centering
  \begin{minipage}{.45\columnwidth}
    \centering
    \includegraphics[width=\columnwidth]{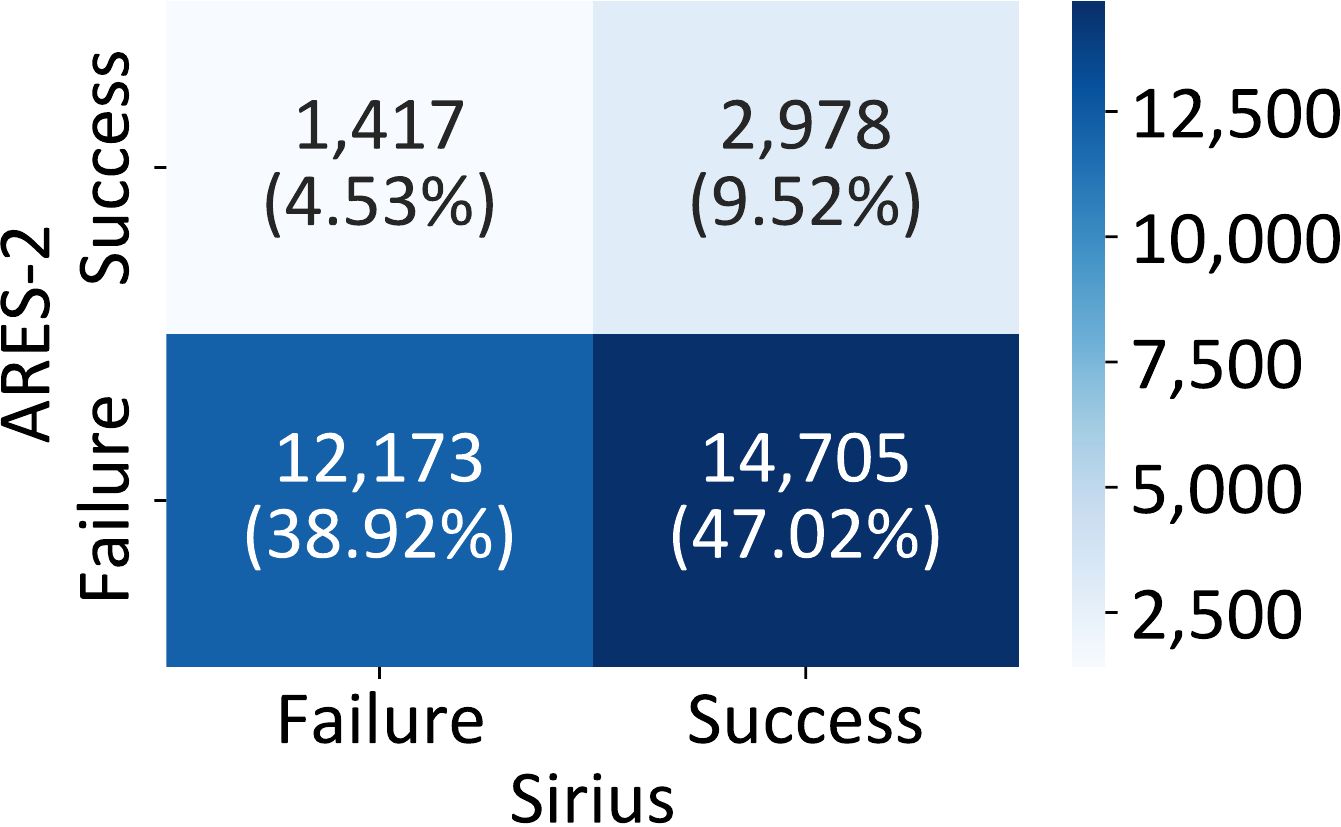}
    \caption{Numbers of success and failure cases between Sirius and ARES-2.}
    \label{fig:cross-table}
  \end{minipage}
  \begin{minipage}{.03\columnwidth}
    \hspace{.3em}
  \end{minipage}
  \begin{minipage}{.47\columnwidth}
    \centering
    \includegraphics[width=\columnwidth]{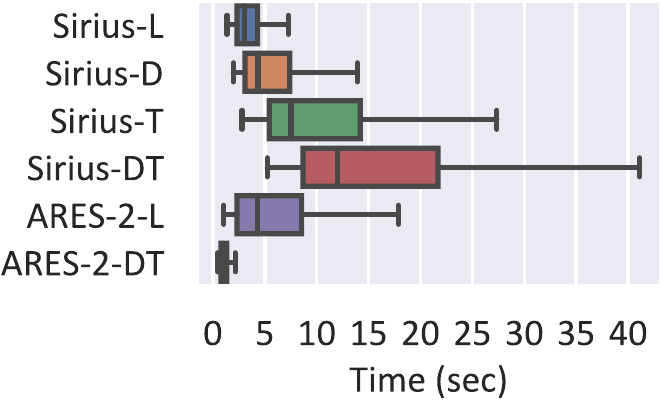}
    \caption{Time spent in repair. L, D, T, and DT denote learning (mining), detection, transformation, and the total of D and T, respectively.}
    \label{fig:time}
  \end{minipage}
\end{figure}

\Cref{fig:cross-table} shows the relation of success and failure cases between Sirius and ARES-2.
Importantly, the number of cases for which Sirius failed but ARES-2 succeeded accounts for only 4.53\% of the total
 while that for which Sirius succeeded but ARES failed accounts for 47.02\%.
This means that Sirius has an advantage over ARES-2 for almost all SEPs.

We sampled 100 Sirius's failure cases to manually examine their causes.
We partitioned the failure cases into four categories on the basis of the size $s$ of each pattern graph:
 small ($s \le 4$); medium ($5 \le s \le 10$); large ($11 \le s \le 20$); and very large ($21 \le s$).
We used stratified sampling with proportional allocation over the four categories.
As a result, we identified the following major causes of repair failures.

\begin{itemize}
  \item \textbf{(1) Incorrect mined SEPs:} This type accounts for 40\% of the total failure cases.
        All the failure cases of this type are due to the lack of examples to learn.
        We excluded future commits from training data as described in \Cref{sec:evaluation-procedure},
         which can lead to insufficiency of training data.
        Patterns mined from fewer training examples can contain changes specific only to the training data, meaning insufficient generalization of the patterns.
        Such patterns can be inapplicable to other code locations, or can introduce unnecessary modifications (specific only to the training data) into test data.
  \item \textbf{(2) Excessive transplantation of AST nodes:} This type accounts for 24\% of the total.
        Sirius transplants AST nodes in units of MTSs.
        MTSs by definition can include AST nodes outside SEPs.
        It means that MTS-based transplantation intrinsically has a risk of excessive transplantation.

        For example, we observed that arguments of method invocations sometimes were unnecessarily modified via MTS-based transplantation.
        Suppose that \textit{getLicense} in the motivating examples requires an argument and the same expression is specified as its argument before and after the change in each example
        (i.e., l.3 in \Cref{fig:code-delta1} is ``... = \textit{app.getLicense(Env.getDefault())};'' and l.3 in \Cref{fig:code-delta2} is ``... = \textit{app.getLicense(env)};'').
        Also, suppose that variable $e$ is specified as the argument of \textit{getLicense} at l.3 in the client (\Cref{fig:client-method}).
        In this situation, the arguments of \textit{getLicense} are uncommon between the examples and thus they are not included in the mined SEP.
        The ground truth of the client code after repair should contain \textit{readLicense(e)}; argument $e$ must not be changed by repair.
        However, when Sirius transplants AST nodes from \astree{W}{new}{ex1} to the client,
         argument \textit{Env.getDefault()} is included in MTS(readLicense) (see \Cref{fig:mts-transplantation1});
         thus, \textit{readLicense(Env.getDefault())} is inserted into the client (i.e., argument $e$ is changed), resulting in a repair failure.
  \item \textbf{(3) Insufficient deletion of AST nodes:} This type accounts for 17\% of the total.
        Currently, our algorithm simply deletes AST nodes corresponding to fgPDG nodes in pattern's old graphs \fgpdg{S}{old}{p}.
        This deletion strategy is sometimes insufficient.
        For example, although a method invocation node \textit{MI} is in \fgpdg{S}{old}{p}, the receiver object of \textit{MI} might not be in \fgpdg{S}{old}{p}.
        In that case, Sirius deletes only the AST node of \textit{MI}, leaving the receiver object in the client code.
        This can lead to unparsable (invalid) code; thus, the result is judged as failures.
  \item \textbf{(4) Incorrect detection results:} This type accounts for 8\% of the total.
        Sometimes, the detector reported more locations than necessary.
        Suppose that there is an SEP that replaces \textit{m1()} with \textit{m2()}, and the client has three invocations of \textit{m1()}.
        Even if the correct change is to replace only two of the three \textit{m1()} with \textit{m2()},
         the detector will report all the three occurrences of \textit{m1()} as the locations to transform, resulting in excessive transformation.
  \item \textbf{(5) Imprecision in AST differencing:} This type accounts for 8\% of the total.
        We compute the AST node mappings between \astree{W}{new}{p} and \astree{W}{old}{p} via a traditional AST differencing algorithm.
        Sometimes, the computed mappings are imprecise, leading to repair failures.
\end{itemize}

It is worth noting that, although it is not a major cause, we observed two cases where variable names were incorrectly determined.
As described in \Cref{sec:label-concretization}, we heuristically determine variable names via mappings between pattern and client graphs.
Some of the mappings are computed on the basis of path similarities in fgPDGs (\Cref{alg:map-pattern-new-to-client}), which can lead to some imprecision.
In our experiment, such imprecision led to incorrect variable names.

Causes (1), (4), and (5) are common problems with Sirius and existing techniques for syntax-based SEPs,
 while (2) and (3) are specific to our algorithm.
These results point out possible directions of future research:
 developing techniques that can learn transformations from much fewer examples (for (1));
 improving the AST transformation algorithm (for (2) and (3));
 and making the detector more intelligent so that we can control locations to transform appropriately (for (4)).

\subsection{RQ2: Efficiency}

\Cref{fig:time} shows the distribution of the time spent on one repair trial (i.e., one iteration in the cross-validation).
We do not plot the individual time of detection and transformation for ARES-2 because they cannot be invoked individually.

Sirius is more time-efficient in learning (mining) while ARES is more time-efficient in detection and transformation.
In terms of practicality, the time efficiency of detection and transformation should be more important because learning can be carried out in advance.

Sirius needs to compute subgraph isomorphism between pattern and client graphs in the detection phase, which takes no small amount of time.
Transformation templates (patterns) inferred by ARES contain the information on how to edit AST nodes, whereas change graphs obtained by Sirius do not indicate AST edit operations.
In our current implementation, Sirius needs to compute AST mappings via dependence graphs and MTSs in the transformation phase, which results in lower efficiency.
Note that because some of the data used in the transformation phase can be computed in advance (e.g., MTSs), there is room for improvement of the time efficiency of the transformer in Sirius.

The median time for applying a systematic edit with Sirius is $\approx$12 seconds.
It would be a more suitable use-case to integrate Sirius into continuous integration environments as a repair bot, where a short response time is not necessarily required, rather than using it in real-time within IDEs during coding tasks.

\subsection{Threats to Validity}
\label{sec:threats-to-validity}

Although the dataset we used in our evaluation contains various OSS projects,
 we cannot further generalize the experimental results to other projects.
However, our transformation algorithm is general-purpose and is not limited to specific domains.
The technique should work well for projects outside the dataset.

We evaluated ARES in two different settings: ARES-2 and ARES-all.
We have not searched for the optimal number of change examples to provide to ARES.
Changing the number of input examples might improve the performance of ARES.
However, it is intrinsically difficult for ARES to infer transformation patterns from a set of code examples with many structural inconsistencies.
Sirius should work better than ARES for fgPDG-based SEPs regardless of the number of input examples.

\section{Related Work}
\label{sec:related-work}

\subsection{Systematic Edit Patterns}

\subsubsection{Change Pattern Mining}

Most existing techniques for mining change patterns are based on syntax-based analyses.
The simplest way for mining SEPs is line/token diff-based analysis\tcite{ammonia:emse2020,precfix:icse2020}.
After anonymizing identifier names that can be uncommon among different methods (e.g., variable names), they find common changes of tokens, resulting in line/token-based SEPs.
To identify the type of each token (e.g., method name or variable), they also leverage AST-based analysis.

SysEdMiner\tcite{sysedminer:msr2017} and C3\tcite{c3:msr2016} use AST diff-based approaches to obtain SEPs.
This type of approach computes the AST edit script for each change example with AST-differencing algorithms\tcite{gumtree:ase2014,change-distiller:tse2007}.
It then finds generalized AST edit operations that are common among examples via frequent itemset mining or clustering (e.g., DBSCAN\tcite{dbscan:kdd1996}).

CPatMiner\tcite{cpatminer:icse2019} is a recently-proposed SEP mining technique.
It detects frequent fgPDG changes as fgPDG-based SEPs.
Compared with traditional syntax-based techniques, it can capture more meaningful changes and is more robust against structural inconsistencies among examples.

While all the above techniques can be used for general purposes, several techniques are specialized for specific domains.
For example, A3\tcite{a3:tse2020} mines API migration patterns from API documents and change examples,
 and Phoenix\tcite{phoenix:fse2019} mines fix patterns for static checker warnings for automated repair.

\subsubsection{Program Transformation}

Program transformation techniques detect common change elements or edit operations among change examples, then generalize them as executable transformations
 that apply change patterns to other locations.

Most transformation techniques analyze syntax-based edit representations to obtain executable transformations.
Precfix\tcite{precfix:icse2020} involves token-level generalization to change examples and infers generic patch templates.
Lase\tcite{lase:icse2013} find common AST edit operations among examples and relate them with dependent code elements to represent edit contexts.
It applies the generalized AST edit operations to the locations having the same contexts.
Rase\tcite{rase:icse2015} (built on Lase) automates a series of systematic editing for refactoring.
VuRLE\tcite{vurle:esroics2017} generates multiple repair templates for each type of vulnerability fixes, which increases the precision and recall of repair.
ARES\tcite{ares:fse2017,Dotzler:phdthesis2018:mtdiff:c3:ares} leverages the specialized transformation templates.
It effectively handles variations among input examples and move-operations, resulting in better accuracy than Lase.

A few techniques\tcite{sydit:pldi2011,sydit:demo:fse2011,genpat:ase2019} can learn a transformation from just one example while most existing tools learn from multiple examples.
Because some types of systematic edits (e.g., bug fixes) do not frequently occur, learning from just one example is an important feature in terms of practicality,
 although they can be imprecise due to fewer examples (e.g., excessive transformation described in \Cref{sec:use-of-existing-techniques}).

The above techniques that find common syntactical edit operations do not work well for fgPDG-based SEPs
 because there are many structural inconsistencies among input examples and client code.

Refazer\tcite{refazer:icse2017} leverages a domain-specific language (DSL) representation to express transformations.
Direct comparison between Refazer and Sirius is difficult because the target programming languages are different.
However, Refazer's DSL does not take into account program dependencies; thus, it would not work well for fgPDG-based SEPs.

There are some techniques specialized for specific domains and use-cases:
 API migrations\tcite{a3:tse2020,libsync:oopsla2010,CocciEvolve:icpc2020,appevolve:issta2019}, fixes of static checker warnings\tcite{phoenix:fse2019,getafix:oopsla:curry-on:2019},
 and real-time recommendations in IDEs\tcite{bluepencil:oopsla2019}.
Of these, A3\tcite{a3:tse2020} and LibSync\tcite{libsync:oopsla2010}, which automate API migrations, leverage dependence graph-based representations of change examples.
They require additional mappings between old and new API methods as their inputs, whereas Sirius does not require them.
Also, their transformation algorithms do not address the difficulties described in \Cref{sec:difficulty-pdg-trans-code-recons};
 thus, we cannot leverage their algorithms to apply fgPDG-based SEPs to other code locations.

\subsection{Test-Driven Program Repair}

There is a large body of work in the research field of automated program repair (APR)\tcite{repair-survey:tse:2019,repair-living-review:techreport:2018,empirical-apr-tools:fse2019,apr:cacm2019}.
Most of them leverage test cases in their repair flows,
 while repair techniques based on SEPs (including Sirius) do not require any test code.

Test-driven APR techniques first localize buggy locations, then generate repair patches.
Typically, fault localization (FL) is carried out by analyzing test outputs\tcite{survey-fault-localization:tse:2016,sapfix:icse-seip:2019}.
After FL, they generate candidate patches by, for example, using predefined templates\tcite{tbar:issta:2019} or machine translation techniques\tcite{coconut:issta2020,cure:icse2021}.
They validate generated patches against the existing test cases to obtain correct (plausible) ones.
To better explore the search space of candidate patches, some techniques leverage machine-learned models\tcite{elixir:icse:2018,prophet:popl2016} or past change patterns\tcite{fixminer:emse2020,hercules:icse:2019,history-driven-repair:saner2016}.
Generated patches that pass existing test cases are often afflicted by overfitting problems.
Many techniques have tried to address the issue via, for example, test case generation\tcite{alleviate-overfit-nopol:emse:2019}.

One of the major practical issues of test-driven repair is the lack of test cases.
Test-driven repair requires test cases for FL and patch validation;
 however, such test cases are not available before bug fixing in realistic situations\tcite{elixir-in-practice:saner2020,ifixr:fse:2019}.
On the other hand, SEP-based repair does not require any test cases and its feasibility is not impaired by this practical issue.

\section{Conclusion}
\label{sec:conclusion}

Developers continually conduct systematic editing, similar but nonidentical changes to many code locations, along with software evolution.
Such repetitive code editing is often error-prone, leading to overlooked (buggy) code locations that require systematic edits.

In this paper, we proposed Sirius, a repair pipeline that automates all processes of mining fgPDG-based SEPs,
 detecting overlooked locations that require systematic edits,
 and repairing the detected code locations.
Sirius leverages our general-purpose program transformation algorithm for applying the same code changes as in fgPDG-based SEPs to other code locations.
There can be many structural inconsistencies among pattern instances and client code;
 therefore, traditional program transformation techniques cannot handle code changes captured via fgPDG-based SEPs well.
Our algorithm overcomes this difficulty by computing local and partial AST node mappings via dependence graphs,
 and transplanting AST substructures in units of MTSs.

We evaluated Sirius with a corpus of OSS consisting of over 80 real-world projects.
The results indicate that Sirius outperformed the state-of-the-art program transformation technique regarding precision, recall, and F1-score.
Sirius achieved good precision and recall for most of the mined patterns, which would be valuable for practical use.

\IEEEtriggeratref{25}

\bibliographystyle{IEEEtran}
\bibliography{IEEEabrv,main}

\end{document}